\documentclass[prb,aps,twocolumn,amsmath,amssymb]{revtex4}
\usepackage{graphicx,bm}
\usepackage{dcolumn}

\begin{document}

\title{Are Quasiparticles and Phonons Identical in Bose--Einstein Condensates?}

\author{Kazumasa Tsutsui,$^1$  Yusuke Kato,$^2$ and Takafumi Kita$^1$}
\affiliation{$^1$Department of Physics, Hokkaido University, Sapporo 060-0810, Japan\\
$^2$Department of Basic Science, The University of Tokyo, 3-8-1 Komaba, Tokyo 153-8902, Japan} %\\

\begin{abstract}
We study an interacting spinless Bose--Einstein condensate to clarify theoretically whether the spectra of its quasiparticles (one-particle excitations) and collective modes (two-particle excitations) are identical, as concluded by Gavoret and Nozi\`eres [Ann. Phys. {\bf 28}, 349 (1964)].
We derive analytic expressions for their first and second moments so as to extend the Bijl--Feynman formula for the peak of the collective-mode spectrum
to its width (inverse lifetime) and also to the one-particle channel.
The obtained formulas indicate that the width of the collective-mode spectrum manifestly vanishes in the long-wavelength limit, 
whereas that of the quasiparticle spectrum apparently remains finite.
We also evaluate the peaks and widths of the two spectra numerically  for a model interaction potential
in terms of the Jastrow wave function optimized by a variational method.
It is thereby found that the width of the quasiparticle spectrum increases towards a constant as the wavenumber decreases. 
This marked difference in the spectral widths implies that the two spectra are distinct.
In particular, the lifetime of the quasiparticles remains finite even in the long-wavelength limit.
\end{abstract}

\maketitle

\section{Introduction}
Gavoret and Nozi\`eres\cite{GN} concluded in 1964 that the spectra of quasiparticles and density-fluctuation modes in an interacting spinless Bose--Einstein condensate are identical by comparing the one- and two-particle Green's functions in their diagrammatic structures of simple perturbation expansions.
The spectra at long wavelengths were also identified as phonons with a linear dispersion, in agreement with previous studies by Bogoliubov \cite{Bogoliubov} and Beliaev\cite{Beliaev}
on the one-particle excitations in the weak-coupling regime.
According to the Gavoret--Nozi\`eres theory,\cite{GN} we may regard collective modes observed by inelastic neutron scattering\cite{VanHove} in superfluid $^4$He\cite{HW61}
as qualitatively identical to the Bogoliubov mode.\cite{Bogoliubov} 
Together with the Hugenholtz--Pines theorem,\cite{HP}
their result has been accepted widely as forming a microscopic foundation for elementary excitations of interacting Bose--Einstein 
condensates.\cite{HM65,MW67,SK74,WG74,text-Pines_Nozieres_vol2,Griffin93}

A question on this was raised recently,\cite{cg-Kita2} however, on the basis of an analysis of the relevant diagrammatic structures
by an alternative self-consistent perturbation expansion,\cite{cg-Kita1,cg-Kita4}
which has a plausible feature of satisfying the Hugenholtz--Pines theorem\cite{HP} and conservation laws\cite{KB,Baym} simultaneously.
It was shown that the single-particle-like line in the two-particle Bethe--Salpeter equation, which is characteristic of condensation,
cannot be identified with the one-particle Green's function
due to extra terms in the ``self-energy'' part,\cite{cg-Kita2} contrary to the conclusion by Gavoret and Nozi\`eres.\cite{GN}
Furthermore, the difference in the self-energies was predicted to manifest itself in the lifetimes of the quasiparticles (one-particle modes) and collective modes (two-particle modes).
To be specific, the collective modes are long-lived with lifetimes that approach infinity in the long-wavelength limit,
whereas the quasiparticles should have much shorter lifetimes as they are bubbling into and out of the condensate dynamically.\cite{cg-Kita3,Tsutsui2}

With this background, we here perform an independent study
as to which of the contradictory conclusions is correct.
The moment method may be suitable for this purpose.
For interacting systems, it is generally practically impossible to calculate dynamical quantities relevant to excitation spectra exactly,
such as the dynamic structure factor\cite{VanHove} $S({\bf q},\omega)$ and  the one-particle spectral function $A({\bf k},\varepsilon)$.
However, we can sometimes obtain reliable results for their moments, which are much easier to handle but nevertheless provide sufficient information
to construct excitation spectra.
Indeed, studying the first moment of $S({\bf q},\omega)$ has proved useful for elucidating the collective-mode spectra 
of such noteworthy systems as superfluid $^4$He\cite{Feynman,FeynmanCohen} and the fractional quantum Hall states.\cite{Girvin} 
We extend this approach up to the second-order moment and also apply it to $A({\bf k},\varepsilon)$
so as to compare the spectral peaks and lifetimes between the one- and two-particle excitations.
This amounts to generalizing the Bijl--Feynman formula\cite{Bijl,Feynman,FeynmanCohen} for the peak of $S({\bf q},\omega)$ to the spectral width 
and also to the single-particle channel.
We subsequently perform a numerical study based on the lowest-order constrained variational (LOCV) method\cite{Pandharipande71,PB73,CHMMPP02,RSAT14} 
to (i) optimize the Jastrow wave function\cite{Jastrow} for the ground state,
(ii) use it to calculate the first and second moments of the quasiparticle and collective-mode spectra numerically by the Monte Carlo method, 
and (iii) compare their peaks and widths for our purpose.

It is worth pointing out that this issue also has direct relevance to the understanding and characterization of Goldstone bosons.\cite{Goldstone61,GSW62,Weinberg96} 
Specifically, Goldstone's theorem was proved in two different manners,\cite{GSW62,Weinberg96} 
which generally have been regarded as identical. 
On the other hand, the first and second proofs in the context of single-component Bose--Einstein condensates 
correspond to the one- and two-particle channels, respectively.\cite{cg-Kita3}
Hence, whether the two excitations are the same or not concerns
the fundamental question of whether the two proofs of Goldstone's theorem are identical or not.

This paper is organized as follows.
Section \ref{MOM} presents the system to be considered, outlines the moment method used for it, and also provides expressions of up to the second moment for 
$S({\bf q},\omega)$ and  $A({\bf k},\varepsilon)$; 
their detailed derivations are given in Appendix.
Section \ref{NC} presents numerical results on the spectral peaks and widths.
A brief summary is given in Sect.\ \ref{Sec:summary}.
We set $\hbar=1$ throughout.

\section{Excitations and Moments}\label{MOM}

We consider a system of $N$ identical spinless bosons with mass $m$ interacting through a two-body potential $U(r)$ in a box of volume ${V}$
with periodic boundary conditions. 
The Hamiltonian is given by
\begin{align}\label{TW_Ham}
\hat H \equiv& \int d^3 r\ \hat\psi^{\dagger}({\bf r})\left(-\frac{\nabla^2}{2m}\right)\hat\psi({\bf r})\notag\\
&+\frac{1}{2}\int d^3 r\int d^3 r'\, U(|{\bf r}-{\bf r}'|)\,\hat\psi^{\dagger}({\bf r})\hat\psi^{\dagger}({\bf r}')\hat\psi({\bf r}')\hat\psi({\bf r})\notag\\
&=\sum_{\bf k}\frac{{\bf k}^2}{2m}\hat c^\dagger_{\bf k}\hat c_{\bf k}+\frac{1}{2V}\sum_{{\bf k}{\bf k}'{\bf q}}U_q\,\hat c^{\dagger}_{{\bf k}+{\bf q}}\hat c^{\dagger}_{{\bf k}'-{\bf q}}\hat c_{{\bf k}'}\hat c_{{\bf k}},
\end{align}
where $\hat\psi^{\dagger}({\bf r})$ and $\hat\psi({\bf r})$ are creation and annihilation operators, respectively, satisfying the Bose commutation relations.\cite{Text-Kita}
The second expression was obtained by expanding the field operator and interaction potential in plane waves as
\begin{align}\label{G_FF}
\hat\psi({\bf r})=&\,\frac{1}{\sqrt{V}}\sum_{\bf k}\hat c_{\bf k}\,e^{i{\bf k}\cdot{\bf r}},
\\
U(r)=&\,\frac{1}{V}\sum_{\bf q}U_q\,e^{i{\bf q}\cdot{\bf r}}
\label{G_FF2},
\end{align}
respectively.

The basic quantities of our interest are the dynamic structure factor $S({\bf q},\omega)$ and one-particle spectral function $A({\bf k},\varepsilon)$ at zero temperature
defined by\cite{Feynman,FeynmanCohen,Girvin,VanHove,FW,Mahan}
\begin{subequations}
\label{AS-def}
\begin{align}
S({\bf q},\omega)\equiv&\,\frac{1}{N}\sum_\nu|\langle\Psi_\nu^{(N)}|\hat\rho_{\bf q}^\dagger|\Psi_0^{(N)}\rangle|^2\delta(\omega-E_\nu^{(N)}+E_0^{(N)}),
\label{S(q,w)-def} \\
A({\bf k},\varepsilon)\equiv&\,\sum_\nu|\langle\Psi_\nu^{(N+1)}|\hat c_{\bf k}^\dagger|\Psi_0^{(N)}\rangle|^2\delta(\varepsilon-E_\nu^{(N+1)}+E_0^{(N)}).
\label{A(k,e)-def}
\end{align}
\end{subequations}
Here $E_\nu^{(N)}$ and $|\Psi_\nu^{(N)}\rangle$ are an eigenvalue of $\hat H$ and its eigenket, respectively,
distinguished by quantum number $\nu$ with $\nu=0$ corresponding to the ground state,
and $\hat\rho_{\bf q}$ is defined by
\begin{align}
\hat\rho_{\bf q}\equiv\sum_{\bf k} \hat c^{\dagger}_{{\bf k}-{\bf q}/2}\hat c_{{\bf k}+{\bf q}/2} =\int d^3 r \,e^{-i{\bf q}\cdot {\bf r}} \hat\psi^\dagger({\bf r})\hat\psi({\bf r}) ,
\label{rho_q^dagger}
\end{align}
satisfying $\hat\rho_{\bf q}^\dagger=\hat\rho_{-{\bf q}}$.
Note that operating $\hat c_{\bf k}^\dagger$ on the ket in Eq.\ (\ref{A(k,e)-def})  increases the particle number by one.

Next, we introduce the moments of $S({\bf q},\omega)$ and $A({\bf k},\varepsilon)$ as
\begin{subequations}
\label{JS_n-def}
\begin{align}
S_{ n}({\bf q})\equiv &\, \int_{-\infty}^{\infty}d\omega \,\omega^n S({\bf q},\omega)  
\notag \\
=&\, N^{-1}\sum_{\nu}
|\langle \Psi_{\nu}^{(N)}|\hat \rho_{{\bf q}}^\dagger|\Psi_0^{(N)}\rangle|^2\left(E_{\nu}^{(N)}-E_0^{(N)}\right)^n,
\label{S_n-def}
\\
A_n({\bf k})\equiv &\, \int_{-\infty}^{\infty}d\varepsilon\, \varepsilon^nA({\bf k},\varepsilon)  
\notag \\
=&\, \sum_{\nu}
|\langle \Psi_{\nu}^{(N+1)}|\hat c_{{\bf k}}^\dagger|\Psi_0^{(N)}\rangle|^2\left(E_{\nu}^{(N+1)}-E_0^{(N)}\right)^n,
\label{A_n-def}
\end{align} 
\end{subequations}
with $n=0,1,2,\cdots$.
It follows from Eq.\ (\ref{S_n-def}) that the quantities
\begin{subequations}
\label{omega_q-domega_q}
\begin{align}
\overline{\omega}_{\bf q}\equiv &\,\frac{S_{ 1}({\bf q})}{S_{ 0}({\bf q})} ,
\label{omega_q}
\\
\overline{\Delta\omega}_{\bf q} \equiv &\, \sqrt{\frac{S_{ 2}({\bf q})}{S_{ 0}({\bf q})}-\left[\frac{S_{ 1}({\bf q})}{S_{ 0}({\bf q})}\right]^2},
\label{domega_q}
\end{align}
\end{subequations}
represent the peak and width of the collective-mode spectrum, respectively.
Indeed, Eqs.\ (\ref{omega_q}) and (\ref{domega_q}) can be regarded as the expectation and standard deviation, respectively, of 
the random variable $x_\nu\equiv E_{\nu}^{(N)}-E_0^{(N)}$ with 
probability $p_\nu\equiv |\langle \Psi_{\nu}^{(N)}|\hat \rho_{{\bf q}}^\dagger|\Psi_0^{(N)}\rangle|^2/NS_{ 0}({\bf q})$.
For the quantities corresponding to Eq.\ (\ref{A_n-def}), 
we should take account of the finite difference 
\begin{align}
\mu\equiv E_{0}^{(N+1)}-E_0^{(N)}
\end{align}
in the ground-state energies due to the addition of a particle. We then
transform $E_{\nu}^{(N+1)}-E_0^{(N)}=E_{\nu}^{(N+1)}-E_0^{(N+1)}+\mu$ to  $E_{\nu}^{(N)}-E_0^{(N)}+\mu$ 
to a close approximation.
We thereby obtain expressions for the peak and width of the quasiparticle spectrum as
\begin{subequations}
\label{epsilon_k-depsilon_k}
\begin{align}
\overline{\varepsilon}_{\bf k}\equiv &\,\frac{A_1({\bf k})}{A_0({\bf k})}-\mu 
\label{epsilon_k}
,\\
\overline{\Delta\varepsilon}_{\bf k} \equiv &\, \sqrt{\frac{A_2({\bf k})}{A_0({\bf k})}-\left[\frac{A_1({\bf k})}{A_0({\bf k})}\right]^2},
\label{depsilon_k}
\end{align}
\end{subequations}
respectively.
Note that the chemical potential $\mu$ is irrelevant to the spectral width.

Next, we express the moments in terms of the ground state alone.
To this end, we consider the Fourier transform of Eq.\ (\ref{S(q,w)-def}) with respect to $\omega$:
\begin{align}
\int_{-\infty}^{\infty}d\omega \,e^{-i\omega t} S({\bf q},\omega)= N^{-1}\langle \Psi_0^{(N)} |e^{i\hat H t} \hat\rho_{{\bf q}} e^{-i\hat H t}\hat\rho_{{\bf q}}^\dagger|\Psi_0^{(N)}\rangle ,
\label{S(q,t)}
\end{align}
where we substituted Eq.\ (\ref{S(q,w)-def}) into the integrand, 
performed integration over $\omega$, 
transformed $e^{-iE_\nu^{(N)} t}|\Psi_\nu^{(N)}\rangle$ to $e^{-i\hat H t}|\Psi_\nu^{(N)}\rangle$,  and used $\sum_{\nu}|\Psi_\nu^{(N)}\rangle\langle\Psi_\nu^{(N)}|=1$.
The inverse transform of Eq.\ (\ref{S(q,t)}) yields the desired expression for $S({\bf q},\omega)$ as
\begin{align}
S({\bf q},\omega)\equiv\frac{N^{-1}}{2\pi}\int_{-\infty}^{\infty} dt\,e^{i\omega t}  \langle \Psi_0 |e^{i\hat H t} \hat\rho_{{\bf q}} e^{-i\hat H t}\hat\rho_{{\bf q}}^\dagger|\Psi_0\rangle 
\label{G_Sk0} ,
\end{align}
where we abbreviated $\Psi_0^{(N)}$ to $\Psi_0$.
Let us substitute Eq. (\ref{G_Sk0}) into the integrand of Eq.\ (\ref{S_n-def}), express $\omega^n e^{i\omega t}$ as $(-id/dt)^n e^{i\omega t}$,  perform $n$ partial integrations 
with respect to $t$, and exchange the order of the integrations over $(\omega,t)$.
We can thereby express $S_{ n}({\bf q})$ alternatively as
\begin{align}\label{G_AIn}
S_{ n}({\bf q})=\left.N^{-1} \left(i\frac{d}{dt}\right)^n\langle \Psi_0 |e^{i\hat H t} \hat\rho_{{\bf q}} e^{-i\hat H t}\hat\rho_{{\bf q}}^\dagger|\Psi_0\rangle\right|_{t\rightarrow 0} .
\end{align}
Hence, we obtain
\begin{subequations}\label{G_Is}
\begin{align}
S_{ 0}({\bf q})&=N^{-1}\langle\Psi_0|\hat\rho_{\bf q}\hat\rho_{\bf q}^\dagger|\Psi_0\rangle\label{G_I0},\\
S_{ 1}({\bf q})&=N^{-1}\langle\Psi_0|[\hat\rho_{\bf q},\hat H]\hat\rho_{\bf q}^\dagger|\Psi_0\rangle\label{G_I1},\\
S_{ 2}({\bf q})&=N^{-1}\langle\Psi_0|[[\hat\rho_{\bf q},\hat H],\hat H]\hat\rho_{\bf q}^\dagger|\Psi_0\rangle
\notag \\
&=N^{-1}\langle\Psi_0|[\hat\rho_{\bf q},\hat H][\hat H,\hat\rho_{\bf q}^\dagger]|\Psi_0\rangle,\label{G_I2}
\end{align}
\end{subequations}
with $[\hat A,\hat B]\equiv \hat A\hat B-\hat B\hat A$.
The second expression of Eq.\ (\ref{G_I2}) may be seen to hold
by using $\langle \Psi_0|\hat H\hat A|\Psi_0\rangle=E_0\langle \Psi_0|\hat A|\Psi_0\rangle=\langle \Psi_0|\hat A\hat H|\Psi_0\rangle$,
where $\hat A=-[\hat\rho_{\bf q},\hat H]\hat\rho_{\bf q}^\dagger$ for the present case.

As shown in Appendix, the commutators in Eq.\ (\ref{G_Is}) can be calculated straightforwardly but rather tediously.
The results are expressible in forms suitable for our Monte Carlo calculations as
\begin{subequations}\label{Moments_second}
\begin{align}
S_{ 0}({\bf q})&=1+n \int d^3 r[g_2({\bf r})-1]e^{i{\bf q}\cdot{\bf r}}+N\delta_{{\bf q}{\bf 0}},
\label{S_0} \\
S_{ 1}({\bf q})&=\frac{ q^2}{2m},\label{S_1}\\
S_{ 2}({\bf q})&= \left(\frac{q^2}{2m}\right)^2-\frac{q^2}{m^2}\left[\frac{1}{3}\nabla^2 g_1({\bf r})\biggr|_{{\bf r}={\bf 0}}+C_{jj}({\bf q})\right],
\label{S_2}
\end{align}
\end{subequations}
where  $n\equiv N/{V}$ is the density of particles, and $g_1({\bf r})$, $g_2({\bf r})$, and $C_{jj}({\bf q})$ are defined in terms of the ground-state wave function $\Psi_0({\bf r}_1,{\bf r}_2,\cdots,{\bf r}_N)$ by
\begin{subequations}
\label{g_21}
\begin{align}
g_1({\bf r})\equiv &\,\prod_{j=1}^N \int d^3 r_{j}\ \Psi_0({\bf r}+{\bf r}_1,{\bf r}_2,\cdots,{\bf r}_N)
\notag \\
&\,\times \Psi_0^*({\bf r}_1,{\bf r}_2,\cdots,{\bf r}_N),
\label{g_1}
\\
g_2({\bf r})\equiv&\,V\frac{N\!-\!1}{N}\prod_{j=2}^N \int d^3 r_{j} |\Psi_0({\bf r}+{\bf r}_2,{\bf r}_2,{\bf r}_{3},\cdots, {\bf r}_{N})|^2,
\label{g_2}
\\
C_{jj}({\bf q})\equiv &\,\frac{N\!-\!1}{4}\prod_{j=1}^N \int d^3 r_{j} e^{-iq\cdot(z_1-z_2)}\left(\frac{\partial}{\partial z_1}-\frac{\partial}{\partial z_1'}\right)
 \notag \\
 &\,\times \left(\frac{\partial}{\partial z_2}-\frac{\partial}{\partial z_2'}\right)
 \Psi_0({\bf r}_1,{\bf r}_2,{\bf r}_3,\cdots,{\bf r}_N)
 \notag \\
 &\,\times
\Psi_0^*({\bf r}_1',{\bf r}_2',{\bf r}_3,\cdots,{\bf r}_N)\biggr|_{{\bf r}_1'={\bf r}_1,{\bf r}_2'={\bf r}_2}
\label{C_jj-def}
\end{align}
\end{subequations}
respectively. 
Function $g_2({\bf r})$ is the pair (or radial) distribution function\cite{Girvin,Mahan,Text-Kita} that obeys the sum rule 
\begin{subequations}
\begin{align}
n \int d^3 r [1-g_2({\bf r})] =1 ,
\label{g_2-sum}
\end{align}
resulting from $S_{ 0}({\bf 0})=N^{-1}\langle\Psi_0|\hat\rho_{\bf 0}\hat\rho_{\bf 0}|\Psi_0\rangle=N$ and Eq.\ (\ref{S_0});
it can also be seen to hold by integrating Eq.\ (\ref{g_2}) over ${\bf r}$ directly.
On the other hand, $g_1({\bf r})$ is essentially the one-particle density matrix\cite{Yang,Mahan,Text-Kita} that satisfies
\begin{align}
g_1({\bf 0})=1
\label{g_1-sum}
\end{align}
\end{subequations}
due to the normalization of $\Psi_0$.
Equation (\ref{S_1}) is known as the $f$-sum rule.\cite{text-Pines_Nozieres_vol1,Mahan}
Finally, Eq.\ (\ref{C_jj-def}) measures a kind of current-current correlation in the ground state, which is expected to be negligible compared with
the other two contributions.

The substitution of Eqs.\ (\ref{S_0}) and (\ref{S_1}) into Eq.\ (\ref{omega_q}) reproduces the Bijl--Feynman formula:\cite{Bijl,Feynman,Girvin,Mahan}
\begin{subequations}
\label{Bijl--Feynman12}
\begin{align}
\overline{\omega}_{\bf q}=\frac{q^2}{2mS_{ 0}({\bf q})},
\label{Bijl--Feynman}
\end{align}
where $S_{ 0}({\bf q}\rightarrow{\bf 0})\rightarrow 0$ when it is continuous at $q=0$, as seen from Eqs.\ (\ref{S_0}) and (\ref{g_2-sum}).
Similarly, Eqs.\ (\ref{domega_q}) and (\ref{Moments_second}) give us the expression
\begin{align}
\overline{\Delta\omega}_{\bf q}=\sqrt{\left\{ -2\frac{g_1''(0)+C_{jj}({\bf q})}{m}
+ [S_{ 0}({\bf q})-1]\overline{\omega}_{\bf q}\right\}\overline{\omega}_{\bf q}},
\label{Bijl--Feynman2}
\end{align}
\end{subequations}
where we evaluated $\nabla^2 g_1({\bf r})\bigr|_{{\bf r}={\bf 0}}$ for isotropic systems by replacing $g_1({\bf r})$ with $g_1(r)$
and noting that $g_1(r\rightarrow 0)$ should behave as $g_1(r)\approx 1-c_2 r^2$ with some constant $c_2>0$.
Equation (\ref{Bijl--Feynman2}) indicates that the width of the collective-mode spectrum vanishes in the long-wavelength limit.
Specifically, $\overline{\Delta\omega}_{\bf q}\propto q^{1/2}$ when $S_{ 0}({\bf q})\propto q$ for $q\rightarrow 0$.

We can also obtain analytic expressions for the moments of $A({\bf k},\varepsilon)$ as detailed in Appendix. 
The results are summarized as
\begin{subequations}\label{Moments_first}
\begin{align}
A_0({\bf k})=&\,1+n\int d^3 r\ g_1({\bf r})e^{-i{\bf k}\cdot{\bf r}},\label{nom_J0}\\
A_1({\bf k})=&\,\frac{k^2}{2m}\left[1+n\int d^3 r\ g_1({\bf r})e^{-i{\bf k}\cdot{\bf r}}\right]
\notag \\
&\,+n\left\{U_0+ \int d^3 r \left[ U(r)g_1({\bf r})+ I_A({\bf r})\right]e^{-i{\bf k}\cdot{\bf r}}\right\},\label{nom_J1}
\\
A_2({\bf k})=&\,\left(\frac{k^2}{2m}\right)^2\left[1+n\int d^3 r\ g_1({\bf r})e^{-i{\bf k}\cdot{\bf r}}\right]\notag\\
&\,+\frac{k^2}{m}n \Bigg\{ U_{0}+ \int d^3 r\, \left[ U(r) g_1({\bf r})\!+\! I_A({\bf r})\right]e^{-i{\bf k}\cdot{\bf r}}\Bigg\}\notag\\
&\,+n \int d^3 r\ [U(r)]^2\left[1+g_1({\bf r})e^{-i{\bf k}\cdot{\bf r}}\right]\notag\\
&\,+n^2\int d^3 r\int d^3 r'\ U(r)U(r')g_2 ({\bf r}-{\bf r}')\notag\\
&\,+n\int d^3 r\left[ 2U(r)I_A({\bf r})+I_B({\bf r})+I_C({\bf r})\right]e^{-i{\bf k}\cdot{\bf r}}.
\label{nom_J2}
\end{align}
\end{subequations}
Here $U_{0}$ is the Fourier coefficient of Eq.\ (\ref{G_FF2}) for $q=0$, and  functions  $I_A$, $I_B$, and $I_C$ are defined by
\begin{subequations}\label{func_SABC}
\begin{align}
I_A({\bf r})\equiv&\,(N\!-\!1)\prod_{j=1}^N\int d^3 r_j\ U(|{\bf r}+{\bf r}_1-{\bf r}_2|)\notag\\
&\,\times\Psi_0({\bf r}+{\bf r}_1,{\bf r}_2,\cdots,{\bf r}_N)\Psi^*_0({\bf r}_1,{\bf r}_2,\cdots,{\bf r}_N),
\label{func_SA} \\
I_B({\bf r})\equiv&\,(N\!-\!1)\prod_{j=1}^N\int d^3 r_j \ U(|{\bf r}_1-{\bf r}_2|)U(|{\bf r}+{\bf r}_1-{\bf r}_2|)\notag\\
&\,\times\Psi_0({\bf r}+{\bf r}_1,{\bf r}_2,\cdots,{\bf r}_N)\Psi^*_0({\bf r}_1,{\bf r}_2,\cdots,{\bf r}_N),
\label{func_SB}\\
I_C({\bf r})\equiv &\,(N\!-\!1)(N\!-\!2)\prod_{j=1}^N\int d^3 r_j\ U(|{\bf r}_1-{\bf r}_2|)\notag\\
&\,\times U(|{\bf r}+{\bf r}_1-{\bf r}_3|)\Psi_0({\bf r}+{\bf r}_1,{\bf r}_2,\cdots,{\bf r}_N) 
\notag \\
&\,\times \Psi^*_0({\bf r}_1,{\bf r}_2,\cdots,{\bf r}_N).
\label{func_SC}
\end{align}
\end{subequations}

Substituting Eq.\ (\ref{Moments_first}) into Eq.\ (\ref{epsilon_k-depsilon_k}), we obtain expressions for the peak and width of the quasiparticle spectrum
as
\begin{subequations}\label{e_k-de_k}
\begin{align}
\overline{\varepsilon}_{\bf k}=&\,\frac{k^2}{2m}\!+\!\frac{n}{A_0({\bf k})}
\left\{\!U_0\!+\! \int d^3 r\! \left[ U(r)g_1({\bf r})\!+\! I_A({\bf r})\right]e^{-i{\bf k}\cdot{\bf r}}\!\right\}\notag \\
&\,-\mu,
\label{e_k}
\end{align}
\begin{align}
&\,\overline{\Delta \varepsilon}_{\bf k}
\notag \\
=&\,\left\{\frac{n}{A_0({\bf k})} \int d^3 r\ [U(r)]^2\left[1+g_1({\bf r})e^{-i{\bf k}\cdot{\bf r}}\right]\right.
\notag\\
&\,+\frac{n^2}{A_0({\bf k})}\int d^3 r\int d^3 r'\ U(r)U(r')g_2 ({\bf r}-{\bf r}')\notag\\
&\,+\frac{n}{A_0({\bf k})}\int d^3 r\left[ 2U(r)I_A({\bf r})+I_B({\bf r})+I_C({\bf r})\right]e^{-i{\bf k}\cdot{\bf r}}
\notag \\
&\,\left.-\frac{n^2}{[A_0({\bf k})]^2}\! \left( U_{0}\!+\! \int d^3 r\bigl[ U(r) g_1({\bf r})\!+\! I_A({\bf r})\bigr]e^{-i{\bf k}\cdot{\bf r}}\!\right)^{\!\!2}\right\}^{\!\!\frac{1}{2}},
\label{de_k}
\end{align}
\end{subequations}
where $U_{0}$ is the Fourier coefficient of Eq.\ (\ref{G_FF2}) for $q=0$, and functions 
$(g_1,g_2)$, $A_0$, and $(I_A,I_B,I_C)$ are defined by Eqs.\ (\ref{g_21}), (\ref{nom_J0}), and  (\ref{func_SABC}), respectively.
Equation (\ref{de_k}) does not vanish manifestly for $k\rightarrow 0$, unlike its corresponding Eq.\ (\ref{Bijl--Feynman2}) for the collective modes,
which indicates that $\overline{\Delta \varepsilon}_{\bf k}$ may remain finite even in the long-wavelength limit.
Our numerical study below for a model interaction potential shows that this is indeed the case.

\section{Numerical Results}\label{NC}

\subsection{Procedure}

We performed numerical calculations of spectral peaks and widths simultaneously for the collective modes and quasiparticles based on
Eqs.\ (\ref{Bijl--Feynman12}) and (\ref{e_k-de_k}), respectively.
The interaction potential we used is given by
\begin{align}
U(r)=\frac{U_0}{8\pi r_0^3}e^{-r/r_0},
\label{U(r)}
\end{align}
where $U_0$ and $r_0$ are positive constants. The corresponding Fourier coefficient in Eq.\ (\ref{G_FF2})
is obtained as $U_q=U_0/(1+r_0^2q^2)^2$.
The fundamental quantities for our purpose are given in Eqs.\ (\ref{g_21}) and (\ref{func_SABC}),
which were evaluated variationally by adopting the Jastrow wave function\cite{Jastrow} 
\begin{align}
\Psi({\bf r}_1,\cdots,{\bf r}_N)=\prod_{i<j}f(|{\bf r}_i-{\bf r}_j|) 
\label{Jastrow}
\end{align}
for the ground state of interacting bosons.
The key function $f(r)$ was determined by the LOCV method,\cite{Pandharipande71,PB73,CHMMPP02,RSAT14}  which provides a reasonable and efficient way 
of constructing the pair function
so as to satisfy the bounday condition $f(r\!\rightarrow\!\infty)\!=\!1$.
This method is outlined as follows. Consider a two-particle scattering problem  
described by the following radial Schr\"odinger equation for the $s$-wave channel:
\begin{align}\label{TW_ue1}
\left[-\frac{1}{2m_{\rm red}}\frac{1}{r}\frac{d^2}{dr^2}r+U(r)\right]\phi(r)=\lambda \phi(r),
\end{align}
where $m_{\rm red}\equiv m/2$, and $\lambda>0$ here acts as the external parameter.
The solution $\phi(r)$ for Eq.\ (\ref{U(r)}) generally monotonically increases from a finite value $\phi(0)>0$ towards its first extremum
$\phi_{\rm max}=\phi(r_{\rm max})$ at a certain point $r_{\rm max}=r_{\rm max}(\lambda)$.
Hence, we solve Eq.\ (\ref{TW_ue1}) numerically for 
$u(r)\equiv r\phi(r)$ with $u(0)=0$ and $u'(0)=1$ up to the first extremum point of $\phi(r)=u(r)/r$ for various values of $\lambda$. 
From the class of solutions, we pick out a single energy $\lambda=\lambda_f$ so that the function
\begin{align}
f(r)\equiv\left\{\begin{array}{ll}\vspace{1mm} \phi(r)/\phi_{\rm max} \,\,&\,\, :\,0\leq r\leq r_{\rm max}(\lambda_f) \\
1  \,\,&\,\, :\, r>r_{\rm max}(\lambda_f)
\end{array}\right. ,
\label{f(r)}
\end{align}
which is continuous up to the first derivative, also satisfies
\begin{align}
4\pi n\int_0^{r_{\rm max}(\lambda_f)} [f(r)]^2 r^2 dr =1 
\label{f-cond}
\end{align}
with $n\equiv N/V$.
Equation (\ref{f-cond}) implies that there is only a single neighbor on average within the radius $r_{\rm max}(\lambda_f)$ around 
each particle,
which consistently justifies the procedure of solving the two-particle scattering problem [Eq.\ (\ref{TW_ue1})] within $r\leq r_{\rm max}(\lambda_f)$.
The resulting $f(r)$ is expanded in plane waves as
\begin{align}
f(|{\bf r}|)=\frac{1}{V}\sum_{{\bf k}}f_{{\bf k}} e^{i{\bf k}\cdot{\bf r}}
\end{align}
so as to obey the periodic boundary conditions, which
is substituted into Eq.\ (\ref{Jastrow}) to construct an approximate ground-state wave function.

We used Eq.\ (\ref{Jastrow}) thereby obtained to evaluate Eqs.\ (\ref{g_21}) and (\ref{func_SABC})
numerically by the variational Monte Carlo method.\cite{MC_text1,Review_Boronat}
For example, Eqs.\  (\ref{g_1}) and (\ref{g_2}) are expressible as
\begin{subequations}
\begin{align}
g_{1}({\bf r})=&\,\frac{1}{N} \sum_{j=1}^N \left<\frac{\Psi({\bf r}_1,\cdots,{\bf r}_j+{\bf r},\cdots,{\bf r}_N)}{\Psi({\bf r}_1,\cdots,{\bf r}_j,\cdots,{\bf r}_N)}\right>,\label{g_1-av}
\\
g_{2}({\bf r})=&\,\frac{1}{nN}\sum_{i\neq j}\langle \delta({\bf r}-{\bf r}_i+{\bf r}_j)\rangle ,
\label{g_2-av}
\end{align}
\end{subequations}
respectively, where $\langle\cdots \rangle$ denotes the expectation with respect to the probability density $|\Psi({\bf r}_1,\cdots,{\bf r}_N)|^2$.
These averages were calculated by standard Monte Carlo procedures.\cite{MC_text1,Review_Boronat}
The delta function in Eq.\ (\ref{g_2-av}) was approximated as
\begin{align}
\delta({\bf r}-{\bf r}_i+{\bf r}_j)\approx\frac{\theta(|{\bf r}_i-{\bf r}_j|-r)-\theta(|{\bf r}_i-{\bf r}_j|-r-\Delta r)}{4\pi r^2\Delta r},
\label{deltafn-approx}
\end{align}
where $\theta(x)$ denotes the step function and $\Delta r>0$. 

It is convenient in practical calculations to know the $s$-wave scattering length $a$
for the potential given by Eq.\ (\ref{U(r)}). 
It is determined as $a=m_{\rm red}{\cal U}(0,0)/2\pi$,\cite{LL-Q} where ${\cal U}(k,k')$ denotes the solution of the integral equation for the $s$-wave channel:
\begin{align}
{\cal U}(k,k')={\cal U}_{0}(k,k')-\int_0^\infty \frac{dk_1}{2\pi^2}k_1^2\frac{ {\cal U}_{0}(k,k_1){\cal U}(k_1,k)}{k_1^2/2m_{\rm red}}
\end{align}
with $ {\cal U}_{0}(k,k')=U_0/[(1+r_0^2k^2+r_0^2k^{\prime 2})^2-4r_0^4k^2k^{\prime 2}]$ for Eq.\ (\ref{U(r)}).\cite{Text-Kita}
We remove $U_0$ in favor of $a$ and choose $a=2r_0$ in the following. 
All the results presented below are given in units of 
\begin{subequations}
\begin{align}
a=1, \hspace{5mm} m=1/2,
\end{align}
with
\begin{align}
r_0=0.5,\hspace{5mm}
n=10^{-3},
\end{align}
\end{subequations}
for which we have $U_0\!\approx\! 57.7465$. 

For reference,  we estimated the total energy by substituting Eq.\ (\ref{Jastrow}) into ${\cal E}\equiv \langle\Psi|\hat H|\Psi\rangle/\langle\Psi|\Psi\rangle$
and evaluating $\langle\Psi|\Psi\rangle$ and the potential-energy term by the variational Monte Carlo method.
We obtained ${\cal E}=1.2941(5)\times10^{-2}$, $1.3778(7)\times10^{-2}$, and $1.4178(8)\times10^{-2}$ for $N=50$, $100$, and $150$, respectively, 
with $20000$ main steps.
Thus, the energy per particle has considerable particle-number dependence in our weak-coupling regime, apparently approaching the value $1.53677\times10^{-2}$
estimated by the two-body cluster estimation:\cite{CHMMPP02}
\begin{align}
\frac{{\cal E}}{N}\approx 2\pi n \lambda \int_0^{r_{\rm max}(\lambda_f)} [f(r)]^2 r^2 dr .
\end{align}
In contrast, quantities relevant to Eqs.\ (\ref{g_21}) and (\ref{func_SABC}) were found to converge rapidly in terms of $N$, as seen  below.

We considered $n_{\rm I}\equiv 400$ independent configurations and took $50000$ presteps for each configuration by using independently generated random numbers.
Afterwards, we took $5000$ and $10000$ main steps  for sampling Eq.\ (\ref{g_2}) and the other functions, respectively, to perform their Monte Carlo integration.
Each point and error bar in all the plots below denote the average and corrected sample standard deviation, respectively, 
of the $n_{\rm I}\equiv 400$ independent configurations; errors were found negligible except for $I_C$.
Equations (\ref{g_1}) and (\ref{func_SABC}) were estimated by choosing ${\bf r}$ along 14 different directions,
$(\pm r,0,0)$, $(0,\pm r,0)$, $(0, 0,\pm r)$, $(\pm r/\sqrt{3},\pm r/\sqrt{3},\pm r/\sqrt{3})$, and averaging the outputs.
We confirmed that varying $\Delta r$ around $\Delta r\sim 10^{-2}$ in Eq.\ (\ref{deltafn-approx}) does not change the results substantially.
The contribution of Eq.\ (\ref{C_jj-def}) to Eq.\ (\ref{S_2}) was  found to be $10^{3}$--$10^{4}$ times smaller than that of $g_1''(0)$
and negligible, as expected.

\subsection{Results}\label{Results}

\begin{figure}[bt]
  \begin{center}
   \includegraphics[width=0.8\linewidth]{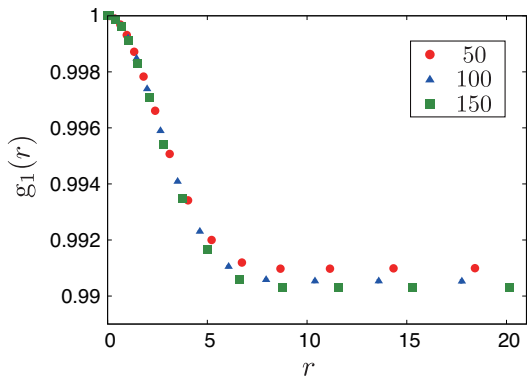}
  \end{center}
  \caption{Plot of $g_1(r)$. \label{fig:rho1}}
\end{figure}
\begin{figure}[bt]
  \begin{center}
   \includegraphics[width=0.8\linewidth]{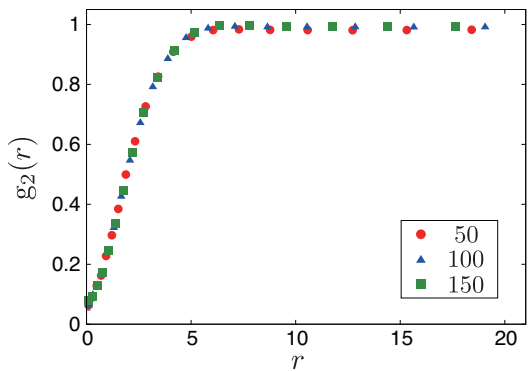}
  \end{center}
  \caption{Plot of $g_2(r)$.\label{fig:g2}}
\end{figure}

\begin{figure}[btp]
  \begin{center}
   \includegraphics[width=0.8\linewidth]{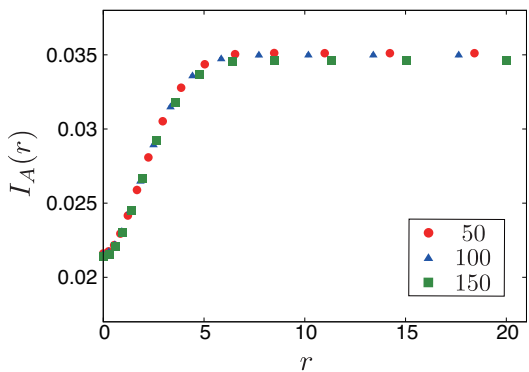}
  \end{center}
  \caption{Plot of $I_A(r)$.\label{fig:nIA}}
\end{figure}

\begin{figure}[btp]
    \begin{center}
   \includegraphics[width=0.8\linewidth]{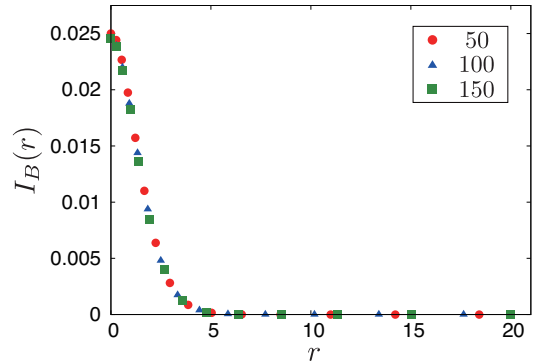}
  \end{center}
  \caption{Plot of $I_B(r)$.\label{fig:nIB}}

\end{figure}

\begin{figure}[btp]
      \begin{center}
   \includegraphics[width=0.8\linewidth]{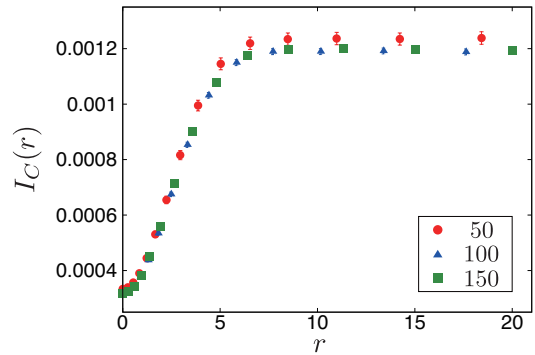}
  \end{center}
  \caption{Plot of $I_C(r)$. \label{fig:nIC}}
\end{figure}

We now present our numerical results for $N=50$, $100$, and $150$.
The basic quantities are the functions in Eqs.\ (\ref{g_21}) and (\ref{func_SABC}).   
Among them, Figs.\ \ref{fig:rho1} and \ref{fig:g2} plot $g_1(r)$ and $g_2(r)$ as functions of $r$, respectively.
$g_1(r)$ starts to gradually decrease from $g_1(0)=1$ towards a finite value $g_1(\infty)>0$,
which is typical of an off-diagonal long-range order\cite{Yang,PO56} with the one-particle density matrix.
The pair distribution function $g_2(r)$ is seen to decrease near the origin, as expected for the repulsive potential in Eq.\ (\ref{U(r)}).
For completeness, we also exhibit the functions of Eq.\ (\ref{func_SABC}) in 
Figs. \ref{fig:nIA}-\ref{fig:nIC}.
They all have a common feature of approaching some constant for $r\gtrsim 5$.


\begin{figure}[tbp]
      \begin{center}
   \includegraphics[width=0.75\linewidth]{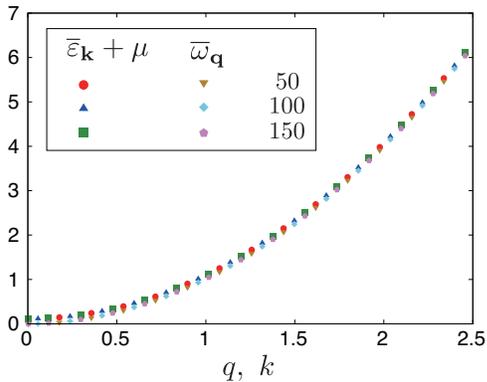}
  \end{center}
  \caption{Spectral peaks of the collective-mode and quasi-particle excitations as functions of momentum. 
For reference, the Bogoliubov theory for the single-particle channel\cite{Bogoliubov} predicts a linear dispersion for $k\ll 0.2242$ with the present parameters.}
  \label{fig:spectrum}
\end{figure}

\begin{figure}[tbp]
      \begin{center}
   \includegraphics[width=0.9\linewidth]{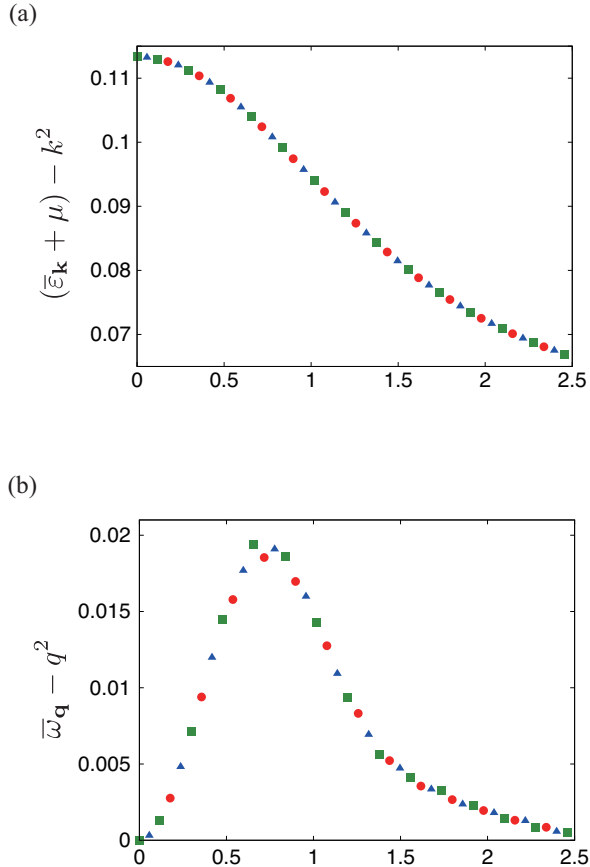}
  \end{center}
  \caption{Deviations from the free-particle spectrum as functions of the momentum for the (a) one-particle excitation and (b) two-particle excitation.}
  \label{fig:deviation}
\end{figure}

Figure \ref{fig:spectrum} shows the spectral peaks of the one- and two-particle excitations as functions of the momentum,
which were
calculated by Eqs.\ (\ref{e_k}) and (\ref{Bijl--Feynman}), respectively.
Both spectra at high momenta have a quadratic dependence on the momentum, as expected.
The two curves are similar to each other even at low momenta except for a finite shift, mainly caused by the presence of $\mu$ in Eq.\ (\ref{e_k}).
Thus, it seems impossible by comparing these curves to conclude definitely whether the two spectra are identical or not,
especially when we consider the difficulty of evaluating $\mu$ with sufficient accuracy for our purpose. 
If we plot the deviations of the two spectra from the free-particle spectrum,  as in Fig.\ \ref{fig:deviation}, however,
we observe a clear difference between the two spectra.
The result strongly indicates that the two spectra are different from each other.

\begin{figure}[tb]
  \begin{center}
   \includegraphics[width=0.75\linewidth]{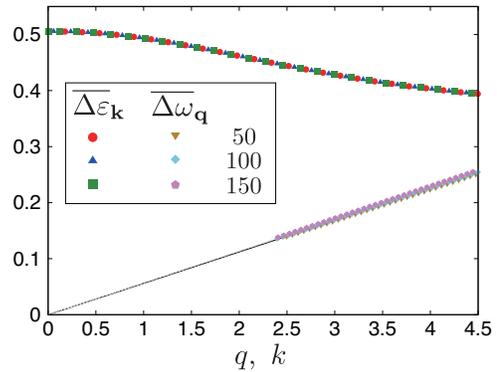}
  \end{center}
  \caption{Half widths of the collective-mode and quasi-particle excitations as functions of momentum. The dotted line is a guide to the eye.}
  \label{fig:life}
\end{figure}

To confirm this point, Fig.\ \ref{fig:life} plots the widths of the two spectra as functions of the momentum, where we can see a marked difference between the two excitations. 
On the one hand, the width of the two-particle spectrum is seen to approach zero linearly at high momenta,
in accordance with Eq.\ (\ref{Bijl--Feynman2}).
On the other hand, that of the one-particle spectrum {\em increases} towards a constant as the momentum approaches 0,
which one may have expected from the analytic formula in Eq.\ (\ref{de_k}) alone.
Unfortunately, it turns out to be impossible to obtain reliable results for $\Delta\omega_{{\bf q}}$ at low momenta 
due to the insufficient numerical accuracy for $r\gtrsim 10$ for the basic functions  in Eqs.\ (\ref{g_21}) and (\ref{func_SABC}). 
Nevertheless, the distinct behaviors of the two widths at high and intermediate momenta
indicate that the two spectra are different from each other.


\section{Summary and Conclusion\label{Sec:summary}}

We have studied the two- and one-particle excitations of a single-component Bose--Einstein condensate at $T=0$ both analytically and numerically,
focusing our attention on the peaks and widths of the spectra.
Their analytic expressions are respectively obtained as Eqs.\ (\ref{Bijl--Feynman12}) and (\ref{e_k-de_k}).
Among them, the widths are given by Eq.\ (\ref{Bijl--Feynman2})  for the two-particle channel and by Eq.\ (\ref{de_k}) for the one-particle channel.
While the former manifestly vanishes in the long-wavelength limit, the latter apparently remains finite at low momenta.
This difference in the spectral widths between the two channels was also confirmed by a numerical study as shown in Fig.\ \ref{fig:life}.
The result implies that, whereas the collective excitations are long-lived, 
 the one-particle excitations are bubbling into and out of the condensate dynamically with finite lifetimes, 
as pointed out previously;\cite{cg-Kita3,Tsutsui2}  these lifetimes
become even shorter at long wavelengths.
In this context, we  mention an experiment on quasiparticle (one-particle) excitations in exciton-polariton condensates.\cite{Utsunomiya}
Although fitted by the Bogoliubov theory, where quasiparticles are predicted to have infinite lifetimes,
the observed one-particle spectrum becomes increasingly broad as the momentum decreases.
These findings suggest the necessity of further theoretical studies on the nature of single-particle excitations in Bose--Einstein condensates.
\\

\noindent
{\bf Acknowledgments}

We are grateful to Dai Hirashima, Naoki Kawashima, and Akiko Masaki-Kato for valuable discussions on our Monte Carlo calculations. 
One of the authors (YK) also thanks K. Hukushima, Y, Sakai and M. Shinozaki
for their instruction on the error-analysis of Monte Carlo simulation.
K.\ T.\ is a JSPS Research Fellow, and this work is supported in part by JSPS KAKENHI Grant Number 15J01505.

\appendix

\section{Derivations of Moments}

Function $S_{ 0}({\bf q})$ can be transformed as follows by substituting Eq.\ (\ref{rho_q^dagger}) in the coordinate representation into Eq.\ (\ref{G_I0}),
moving the field operators into a normal order,\cite{FW} and making the change of integration variables ${\bf r}_1\rightarrow {\bf r}\equiv{\bf r}_1-{\bf r}_2$:
\begin{align}
S_{ 0}({\bf q})=&\,\frac{1}{N}\int \! d^3 r_1\int\! d^3 r_2\!\left[ \delta({\bf r}_1\!-\!{\bf r}_2) \langle\Psi_0| \hat\psi^\dagger({\bf r}_1)\hat\psi({\bf r}_2)|\Psi_0\rangle
\right.
\notag \\
&\,\left.
+ \langle\Psi_0| \hat\psi^\dagger({\bf r}_1)\hat\psi^\dagger({\bf r}_2)\hat\psi({\bf r}_2)\hat\psi({\bf r}_1)|\Psi_0\rangle\right]
e^{-i{\bf q}\cdot({\bf r}_1-{\bf r}_2)}
\notag \\
=&\,1+\frac{N}{V}\int d^3 r  \,g_2({\bf r})\, e^{-i{\bf q}\cdot{\bf r}},
\label{S_0-3}
\end{align}
where $g_2({\bf r})$ is the pair distribution function for homogeneous systems at zero temperature defined by
\begin{align}
g_2({\bf r})\equiv &\,\frac{V}{N^2}\int d^3 r_2\langle\Psi_0| \hat\psi^\dagger({\bf r}+{\bf r}_2)\hat\psi^\dagger({\bf r}_2)\hat\psi({\bf r}_2)
\notag \\
&\,\times \hat\psi({\bf r}+{\bf r}_2)|\Psi_0\rangle .
\label{g_2-0}
\end{align}
Equation (\ref{g_2-0}) is expressible as Eq.\ (\ref{g_2}).\cite{Text-Kita} 
Hence, Eq.\ (\ref{S_0-3}) is identical to Eq.\ (\ref{S_0}).

Next, we focus on the first moment in Eq.\ (\ref{G_I1}).
Assuming the time-reversal symmetry $S_{ 1}({\bf q})=S_{ 1}(-{\bf q})$, 
we express $S_{ 1}({\bf q})=[S_{ 1}({\bf q})+S_{ 1}(-{\bf q})]/2$, 
substitute Eq.\ (\ref{G_I1}) into its right-hand side, 
and use $\hat\rho_{-{\bf q}}=\hat\rho_{{\bf q}}^\dagger$ and $\langle \Psi_0|\hat H\hat A|\Psi_0\rangle=\langle \Psi_0|\hat A\hat H|\Psi_0\rangle$
with $\hat A\equiv -\hat\rho_{{\bf q}}^\dagger\hat\rho_{{\bf q}}$ for the present case.
We thereby obtain
\begin{align}\label{G_I12}
S_{ 1}({\bf q})=\frac{1}{2N}\langle\Psi_0|[[\hat\rho_{\bf q},\hat H],\hat\rho^{\dagger}_{\bf q}]|\Psi_0\rangle.
\end{align}
The commutator $[\hat\rho_{\bf q},\hat H]$ can be calculated easily by omitting from Eq.\ (\ref{TW_Ham}) the interaction part  that commutes with
$\hat\rho_{\bf q}$, as seen easily from the coordinate expressions 
of Eqs.\ (\ref{TW_Ham}) and (\ref{rho_q^dagger}). 
We obtain
\begin{align}\label{G_Jq}
[\hat\rho_{\bf q},\hat H]&=\frac{{\bf q}}{m}\cdot\sum_{{\bf k}}{\bf k}\ a_{{\bf k}-{\bf q}/2}^\dagger\hat c_{{\bf k}+{\bf q}/2},
\end{align}
which denotes the current operator.\cite{text-Pines_Nozieres_vol1}
Using Eq.\ (\ref{G_Jq}) in Eq.\ (\ref{G_I12}) and calculating the commutator, we obtain Eq.\ (\ref{S_1}).
It is worth noting for later purposes that the condition $S_{ 1}({\bf q})=S_{ 1}(-{\bf q})$ used above is expressible as
\begin{align}
&\,
\frac{{\bf q}}{m}\cdot \int d^3 r_1  \int d^3 r_2 
\, e^{-i{\bf q}\cdot({\bf r}_1-{\bf r}_2)}\frac{{\bm\nabla}_1-{\bm\nabla}_1'}{2i}
\notag \\
&\,\times
 \langle\Psi_0| \hat\psi^\dagger({\bf r}_1')\hat\psi^\dagger({\bf r}_2)\hat\psi({\bf r}_2)\hat\psi({\bf r}_1)|\Psi_0\rangle\biggr|_{{\bf r}_1'={\bf r}_1} =0 ,
\end{align}
which vanishes for any $\Psi_0$ with no net momentum.

For the second moment $S_{ 2}({\bf q})$, we substitute Eq.\ (\ref{G_Jq}) into Eq.\ (\ref{G_I2}), arrange the field operators into a normal order,
and make the change of variables ${\bf k}-{\bf q}/2\rightarrow{\bf k}$ to obtain
\begin{align}\label{G_I21}
S_{ 2}({\bf q})
=&\,\sum_{\mu,\nu}\frac{q_\mu q_\nu}{m^2 N}\sum_{{\bf k}}\!\left(k_{\mu}+\frac{q_\mu}{2}\right)\!\!\left(k_{\nu}+\frac{q_\nu}{2}\right)\!
\langle\Psi_0|\hat c^\dagger_{{\bf k}}\hat c_{{\bf k}}|\Psi_0\rangle
\notag\\
&\, +\sum_{\mu,\nu=x,y,z}\frac{q_\mu q_\nu}{m^2 N}\sum_{{\bf k}{\bf k}'}k_{\mu}k_{\nu}'
\langle\Psi_0|\hat c^\dagger_{{\bf k}-{\bf q}/2}\hat c^\dagger_{{\bf k}'+{\bf q}/2}
\notag \\
&\,\times \hat c_{{\bf k}+{\bf q}/2}\hat c_{{\bf k}'-{\bf q}/2}|\Psi_0\rangle  .
\end{align}
We further substitute the inverse  transform of Eq.\ (\ref{G_FF}), write $k_\mu e^{-i{\bf k}\cdot({\bf r}_1-{\bf r}_1')}=(\nabla_{1\mu}'-\nabla_{1\mu})e^{-i{\bf k}\cdot({\bf r}_1-{\bf r}_1')}/2i$, etc., perform partial integrations with respect to the momentum operators, and carry out summations over $({\bf k},{\bf k}')$.
We thereby obtain 
\begin{align}\label{G_I22}
S_{ 2}({\bf q})=\left(\frac{q^2}{2m}\right)^2-\frac{q^2}{m^2}\left[\frac{1}{3}\nabla^2 g_1({\bf r})\bigr|_{{\bf r}={\bf 0}}+C_{jj}({\bf q})\right],
\end{align}
where functions $g_1({\bf r})$ and $C_{jj}({\bf q})$ are defined by
\begin{align}\label{g_1-def}
g_1({\bf r})\equiv \frac{1}{N}\int d^3 r_1\langle\Psi_0|\hat\psi^{\dagger}({\bf r}_1)\hat\psi({{\bf r}+{\bf r}_1})|\Psi_0\rangle .
\end{align}
\begin{align}
&\,
C_{jj}({\bf q})
\notag \\
\equiv &\frac{1}{4N} \int d^3 r_1\int d^3 r_2
e^{-i q\cdot(z_1-z_2)} ({\bm \nabla}_{1}\!-\!{\bm \nabla}_{1}')_z({\bm \nabla}_{2}\!-\!{\bm \nabla}_{2}')_z
\notag \\
&\times 
\langle \Psi_0|\hat\psi^\dagger({\bf r}_1')\hat\psi^\dagger({\bf r}_2')\hat\psi({\bf r}_2)\hat\psi({\bf r}_1)|\Psi_0\rangle\biggr|_{{\bf r}_1'={\bf r}_1,{\bf r}_2'={\bf r}_2}.
\label{C_jj}
\end{align}
In deriving Eq.\ (\ref{C_jj}), we chose ${\bf q}$ along the $z$ axis without loss of generality.
Equations (\ref{g_1-def}) and (\ref{C_jj}) are expressible as  Eqs.\ (\ref{g_1}) and (\ref{C_jj-def}), respectively.\cite{Text-Kita}
Hence, Eq.\ (\ref{G_I22}) is identical to Eq.\ (\ref{S_2}).

Now, we consider $A_n({\bf k})$, which for $n\leq 2$ can be obtained as follows from Eq.\ (\ref{G_Is})
by the replacement $\hat\rho_{{\bf q}}\rightarrow N^{1/2} \hat c_{\bf k}$:
\begin{subequations}\label{JJA_tes}
\begin{align}
A_0({\bf k})&\equiv\langle\Psi_0|\hat c_{\bf k}\hat c^\dagger_{\bf k}|\Psi_0\rangle,\label{G_J0}\\
A_1({\bf k})&\equiv\langle\Psi_0|[\hat c_{\bf k},\hat H] \hat c^\dagger_{\bf k}|\Psi_0\rangle,\label{G_J1}\\
A_2({\bf k})&\equiv\langle\Psi_0|[\hat c_{\bf k},\hat H][\hat H,\hat c^\dagger_{\bf k}]|\Psi_0\rangle.\label{GJ2_test}
\end{align}
\end{subequations}
We transform them into forms suitable for importance sampling.

Writing $\hat c_{\bf k}\hat c_{\bf k}^\dagger=1+\hat c_{\bf k}^\dagger\hat c_{\bf k}$ in Eq.\ (\ref{G_J0}) and 
substituting the inverse transform of Eq.\ (\ref{G_FF}), one can easily show that $A_0({\bf k})$ is expressible as
Eq.\ (\ref{nom_J0}) in terms of $g_1({\bf r})$ defined by Eq.\ (\ref{g_1-def}).

For $A_1({\bf k})$, the commutator of $\hat c_{\bf k}$ with Eq.\ (\ref{TW_Ham}) is obtained as
\begin{align}
[\hat c_{\bf k},\hat H]=\frac{k^2}{2m}\hat c_{\bf k}+\frac{1}{V}\sum_{{\bf k}'{\bf q}}U_q\hat c_{{\bf k}'-{\bf q}}^\dagger\hat c_{{\bf k}'}\hat c_{{\bf k}-{\bf q}},
\label{G_cr1}
\end{align}
We use it in Eq.\ (\ref{G_J1}),  arrange the field operators into a normal order,\cite{FW}
and substitute the inverse transform of Eq.\ (\ref{G_FF}).
We thereby obtain
\begin{align}
A_1({\bf k})
=&\,\frac{{\bf k}^2}{2m}\left[1+ n  \int d^3 r g_1({\bf r})\ e^{-i{\bf k}\cdot{\bf r}}\right]+nU_0 
\notag\\
&\,+n \int d^3 r \,U(r) g_1({\bf r}) \,e^{-i{\bf k}\cdot{\bf r}}
\notag\\
&\,+\frac{1}{V}\int d^3 r_1\int d^3 r_1'\int d^3 r_2\ U(|{\bf r}_1-{\bf r}_2|)\notag\\
&\,\times\!\langle\Psi_0|\hat\psi^\dagger({{\bf r}_1'})\hat\psi^\dagger({{\bf r}_2})\hat\psi({{\bf r}_2})\hat\psi({{\bf r}_1})|\Psi_0\rangle e^{-i{\bf k}\cdot({\bf r}_1-{\bf r}_1')}
\label{A_1(k)},
\end{align}
where $g_1({\bf r})$ is defined by Eq.\ (\ref{g_1-def}) and $U_0$ is the Fourier coefficient in Eq.\ (\ref{G_FF2}) for $q=0$.
We now introduce the function
\begin{align}
I_A({\bf r})\equiv&\,\frac{1}{N}\int d^3 r_1' \int d^3 r_2\ U(|{\bf r}+{\bf r}_1'-{\bf r}_2|)\notag\\
&\,\times\langle\Psi_0|\hat\psi^\dagger({{\bf r}_1'})\hat\psi^\dagger({{\bf r}_2})\hat\psi({{\bf r}_2})\hat\psi({{\bf r}_1'+{\bf r}})|\Psi_0\rangle,
\label{I_A0}
\end{align}
which is identical to Eq.\ (\ref{func_SA}).\cite{Text-Kita}
Let us express the last term of Eq.\ (\ref{A_1(k)}) in terms of Eq.\ (\ref{I_A0}) and make the change of integration variables
${\bf r}_1\rightarrow {\bf r}\equiv {\bf r}_1-{\bf r}_1'$.
We thereby obtain Eq.\ (\ref{nom_J1}).

Finally, we substitute Eq.\ (\ref{G_cr1}) and its Hermitian conjugate into Eq.\ (\ref{GJ2_test}). Transforming the resulting expression similarly,
we obtain
\begin{align}
&\,A_2({\bf k})\notag\\
=&\left(\frac{k^2}{2m}\right)^2\left[1+n\int d^3 r\ g_1({\bf r})e^{-i{\bf k}\cdot{\bf r}}\right]\notag\\
&+2\frac{k^2}{2m}n\left\{U_{0}+\int d^3 r\ \left[U(r)g_1({\bf r})+ I_A({\bf r})\right]e^{-i{\bf k}\cdot{\bf r}}\right\}\notag\\
&+\frac{1}{V}\int d^3 r_1\int d^3 r_1'\int d^3 r_2\int d^3 r_2'\ U(|{\bf r}_1-{\bf r}_2|)\notag\\
&\times U(|{\bf r}_1'\!-\!{\bf r}_2'|)\bigl[\delta({\bf r}_1\!-\!{\bf r}_1')\delta({\bf r}_2\!-\!{\bf r}_2')\langle\Psi_0|\hat\psi^\dagger({{\bf r}_2})\hat\psi({{\bf r}_2})|\Psi_0\rangle\notag\\
&+\delta({\bf r}_1'-{\bf r}_2)\delta({\bf r}_1-{\bf r}_2')\langle\Psi_0|\hat\psi^\dagger({{\bf r}_1'})\hat\psi({{\bf r}_1})|\Psi_0\rangle\notag\\
&+\delta({\bf r}_1-{\bf r}_1')\langle\Psi_0|\hat\psi^\dagger({{\bf r}_2})\hat\psi^\dagger({{\bf r}_2'})\hat\psi({{\bf r}_2'})\hat\psi({{\bf r}_2})|\Psi_0\rangle\notag\\
&+\delta({\bf r}_1-{\bf r}_2')\langle\Psi_0|\hat\psi^\dagger({{\bf r}_1'})\hat\psi^\dagger({{\bf r}_2})\hat\psi({{\bf r}_2})\hat\psi({{\bf r}_1})|\Psi_0\rangle\notag\\
&+\delta({\bf r}_1'-{\bf r}_2)\langle\Psi_0|\hat\psi^\dagger({{\bf r}_1'})\hat\psi^\dagger({{\bf r}_2})\hat\psi({{\bf r}_2})\hat\psi({{\bf r}_1})|\Psi_0\rangle\notag\\
&+\delta({\bf r}_2-{\bf r}_2')\langle\Psi_0|\hat\psi^\dagger({{\bf r}_1'})\hat\psi^\dagger({{\bf r}_2})\hat\psi({{\bf r}_2})\hat\psi({{\bf r}_1})|\Psi_0\rangle\notag\\
&+\langle\Psi_0|\hat\psi^\dagger({{\bf r}_1'})\hat\psi^\dagger({{\bf r}_2})\hat\psi^\dagger({{\bf r}_2'})\hat\psi({{\bf r}_2'})\hat\psi({{\bf r}_2})\hat\psi({{\bf r}_1})|\Psi_0\rangle
\bigr] 
\notag \\
&\,\times e^{-i{\bf k}\cdot({\bf r}_1-{\bf r}_1')}.
\label{A_2-App}
\end{align}
We now introduce the functions
\begin{subequations}
\label{I_BC}
\begin{align}
I_B({\bf r})\equiv &\, \frac{1}{N}\int d^3 r_1'\int d^3 r_2\ U(|{\bf r}+{\bf r}_1'-{\bf r}_2|)U(|{\bf r}_1'-{\bf r}_2|)\notag\\
&\,\times\langle\Psi_0|\hat\psi^\dagger({{\bf r}_1'})\hat\psi^\dagger({{\bf r}_2})\hat\psi({{\bf r}_2})\hat\psi({{\bf r}+{\bf r}_1'})|\Psi_0\rangle,\\
I_C({\bf r})\equiv&\, \frac{1}{N}\int d^3 r_1'\int d^3 r_2\int d^3 r_2'\ U(|{\bf r}+{\bf r}_1'-{\bf r}_2|)
\notag\\
&\,\times U(|{\bf r}_1'-{\bf r}_2'|)\langle\Psi_0|\hat\psi^\dagger({{\bf r}_1'})\hat\psi^\dagger({{\bf r}_2})\hat\psi^\dagger({{\bf r}_2'})
\notag\\
&\,\times\hat\psi({{\bf r}_2'})\hat\psi({{\bf r}_2})\hat\psi({{\bf r}+{\bf r}_1'})|\Psi_0\rangle,
\end{align}
\end{subequations}
which are identical to Eqs.\ (\ref{func_SB}) and (\ref{func_SC}), respectively.\cite{Text-Kita}
We now express Eq.\ (\ref{A_2-App}) by using Eqs.\ (\ref{g_2-0}), (\ref{g_1-def}), (\ref{I_A0}), and (\ref{I_BC}).
We thereby obtain Eq.\ (\ref{nom_J2}).


\begin{thebibliography}{99}
\bibitem{GN}
J. Gavoret and P. Nozi\`eres, Ann. Phys. {\bf 28}, 349 (1964).
\bibitem{Bogoliubov}
N. N. Bogoliubov, J. Phys. (USSR) {\bf 11}, 23 (1947).
\bibitem{Beliaev}S. T. Beliaev,
Zh. Eksp. Teor. Fiz. {\bf 34}, 433  (1958)
[Sov. Phys. JETP {\bf 7}, 299 (1958)].
\bibitem{VanHove}
L. Van Hove, Phys. Rev. {\bf 95}, 249 (1954).
\bibitem{HW61}
A. D. B. Woods and R. A. Cowley, Rep. Prog. Phys. {\bf 36}, 1135 (1973).
\bibitem{HP}
N. M. Hugenholtz and D. Pines, Phys. Rev. {\bf 116}, 489 (1959).
\bibitem{HM65}
P. C. Hohenberg and P. C. Martin, Ann. Phys. (N.Y.) {\bf 34}, 291 (1965).
\bibitem{MW67}
S. K. Ma and C. W. Woo, Phys. Rev. {\bf 159}, 165 (1967).
\bibitem{SK74}
P. Sz\'epfalusy and I. Kondor, Ann. Phys. (N.Y.) {\bf 82}, 1 (1974). 
\bibitem{WG74}
V. K. Wong and H. Gould, Ann. Phys. (N.Y.) {\bf 83}, 252 (1974). 
\bibitem{text-Pines_Nozieres_vol2}
P. Nozi\`eres and D. Pines, {\it The Theory of Quantum Liquids} (W. A. Benjamin, New York, 1990) Vol. II.
\bibitem{Griffin93}
A. Griffin, {\em Excitations in a Bose-Condensed Liquid} (Cambridge University Press, Cambridge, 1993).
%\bibitem{Goldstone}
%J. Goldstone, A. Salam, S. Weinberg,\ Phys. Rev. {\bf 127}, 965 (1962).
%\bibitem{Weinberg}
%S. Weinberg,\ {\it The Quantum Theory of Fields}, (Cambridge University Press, 1995).
\bibitem{cg-Kita2}
T. Kita, Phys. Rev. B {\bf 81}, 214513 (2010).
\bibitem{cg-Kita1}
T. Kita, Phys. Rev. B {\bf 80}, 214502 (2009)
\bibitem{cg-Kita4}
T. Kita,  J. Phys. Soc. Jpn. {\bf 83}, 064005 (2014).
%\bibitem{LW}
%J. M. Luttinger, J. C. Ward,\ Phys. Rev. {\bf 118}, 1417 (1960).
\bibitem{KB}
L. P. Kadanoff and G. Baym, {\it Quantum Statistical Mechanics} (Benjamin, New York, 1962).
\bibitem{Baym} 
G. Baym, Phys. Rev. {\bf 127}, 1391 (1962).
\bibitem{cg-Kita3} 
T. Kita,  J. Phys. Soc. Jpn. {\bf 80}, 084606 (2011).
\bibitem{Tsutsui2}
K. Tsutsui and T. Kita,  J. Phys. Soc. Jpn. {\bf 83}, 033001  (2014).
\bibitem{Feynman}
R. P. Feynman, Phys. Rev. {\bf 94}, 262 (1954).
\bibitem{FeynmanCohen}
R. P. Feynman and M. Cohen, Phys. Rev. {\bf 102}, 1189 (1956).
\bibitem{Girvin}
S. M. Girvin, A. H. MacDonald, and P. M. Platzman, Phys. Rev. B {\bf 33}, 2481 (1986).
\bibitem{Bijl}
A. Bijl, Physica {\bf 7}, 869 (1940).
\bibitem{Pandharipande71}V. R. Pandharipande, Nucl. Phys. A {\bf 178}, 123 (1971).
\bibitem{PB73}V. R. Pandharipande and H. A. Bethe, Phys. Rev. C {\bf 7}, 1312 (1973).
\bibitem{CHMMPP02}S. Cowell, H. Heiselberg, I. E. Mazets, J. Morales, V. R. Pandharipande, and C. J. Pethick, Phys. Rev. Lett. {\bf 88}, 210403 (2002).
\bibitem{RSAT14}M. Rossi, L. Salasnich, F. Ancilotto, and F. Toigo, Phys. Rev. A {\bf 89}, 041602 (2014). 
\bibitem{Jastrow}
R. Jastrow, Phys. Rev. {\bf 98}, 1479 (1955).
\bibitem{Goldstone61}
J. Goldstone, Nuovo Cimento {\bf 19}, 154 (1961).
\bibitem{GSW62}
 J.\ Goldstone, A. Salam, and S. Weinberg: Phys. Rev. {\bf 127}, 965 (1962).
\bibitem{Weinberg96}
S. Weinberg, {\em The Quantum Theory of Fields II} (Cambridge University Press, Cambridge, U.K., 1996).
\bibitem{Text-Kita}
T. Kita, {\it Statistical Mechanics of Superconductivity} (Springer, Tokyo, 2015).
\bibitem{FW}
A. L. Fetter and J. D. Walecka, {\it Quantum Theory of Many-Particle Systems} (McGraw-Hill, New York, 1971).
\bibitem{Mahan}
G. D. Mahan, {\it Many-Particle Physics} (Kluwer Academic / Plenum, New York, 2000) 3rd ed.
\bibitem{Yang}
C. N. Yang, Rev. Mod. Phys. {\bf 34}, 694 (1962).
\bibitem{text-Pines_Nozieres_vol1}
D. Pines and P. Nozi\`eres, {\it The Theory of Quantum Liquids} (W. A. Benjamin, New York, 1966) Vol. I.

\bibitem{MC_text1}
B. L. Hammond, W. A. Lester, Jr., and P. J. Reynolds, {\it Monte Carlo Methods in Ab Initio Quantum Chemistry}, (World Scientific, Singapore, 1994).
\bibitem{Review_Boronat}
J. Boronat, in {\it Microscopic  Approaches to Quantum Liquids in Confined Geometries},
 ed.\
E. Krotscheck and J.Navarro (World Scientific, Singapore, 2002) Chap. 2.
\bibitem{LL-Q}
See, for example, L. D. Landau and E. M. Lifshitz, {\it Quantum Mechanics} (Pergamon, Oxford, 1989), 3rd ed., Eq.\ (132.9).
\bibitem{PO56}
O. Penrose and L. Onsager, Phys. Rev. {\bf 104}, 576 (1956).
\bibitem{Utsunomiya}
S. Utsunomiya, L. Tian, G. Roumpos, C. W. Lai, N. Kumada, T. Fujisawa, M. Kuwata-Gonokami, A. L\"offler, S. H\"ofling, A. Forchel, and Y. Yamamoto, Nat. Phys. {\bf 4}, 700 (2008).

\end{thebibliography}
\end{document}